\documentclass[%
 superscriptaddress,
 reprint,
 showpacs,preprintnumbers,
 nofootinbib,
 amsmath,amssymb,
 aps,
 prl, 
 floatfix,
]{revtex4-1}
\usepackage{graphicx}
\usepackage{xcolor}
\usepackage{subcaption}
\usepackage{array,url}
\usepackage{numprint}
\usepackage{hyperref}
\hypersetup{
    colorlinks=true,
    linkcolor=blue,
    filecolor=blue,
    urlcolor=blue,
    pdftitle={nue_numu_uncertainties},
    }

\usepackage{amssymb,amsmath,amstext,amsthm,amsfonts,nicefrac}
\usepackage{cleveref}

\usepackage{tabularx, makecell}
\usepackage{booktabs} 
\usepackage{xspace}

\usepackage{hyperref} 
\usepackage{ulem}
\usepackage{multirow}
\usepackage{amsmath}

\newcommand\parbar[1]{\overset{\textbf{\fontsize{2pt}{2pt}\selectfont(---)}}{#1}}

\newcommand{\nue}                {$\nu_{e}$\xspace}
\newcommand{\nuebar}           {$\bar{\nu}_{e}$\xspace}
\newcommand{\numubar}        {$\bar{\nu}_{\mu}$\xspace}
\newcommand{\numu}             {$\nu_{\mu}$\xspace}

\RequirePackage{xspace}

\newcommand{\affiliationETHZ}{ETH Zurich, Institute for Particle physics and Astrophysics, CH-8093 Zurich, Switzerland.}
\newcommand{\affiliationCERN}{European Organization for Nuclear Research (CERN), 1211 Geneva 23, Switzerland.}
\newcommand{\affiliationGhent}{Theoretical Physics Department, Fermilab, Batavia IL 60510, USA.}
\newcommand{\affiliationLancaster}{Lancaster University, Physics Department, Lancaster, United Kingdom.}
\newcommand{\affiliationRHUL}{Royal Holloway University of London, Department of Physics, Egham, Surrey, United Kingdom.}
\newcommand{\affiliationCEA}{IRFU, CEA, Université Paris-Saclay, F-91191 Gif-sur-Yvette, France.}

\def\nub        {\ensuremath{\overline{\nu}}\xspace}
\def\nue        {\ensuremath{\nu_e}\xspace}
\def\nueb       {\ensuremath{\nub_e}\xspace}
\def\num        {\ensuremath{\nu_\mu}\xspace}
\def\numb       {\ensuremath{\nub_\mu}\xspace}

\usepackage{lineno}

\definecolor{LightCyan}{rgb}{0.88,1,1}
\usepackage{xcolor,colortbl}

\begin{document}

	\title{Uncertainties on the $\parbar{\nu}_e$/$\parbar{\nu}_{\mu}$ and $\nu_e$/$\bar{\nu}_e$ cross-section ratio from the modelling of nuclear effects at 0.2 to 1.2 GeV neutrino energies and their impact on neutrino oscillation experiments}

%

    \author{T.~Dieminger}
    \email[Contact e-mail: ]{tilld@ethz.ch}
    \affiliation{\affiliationETHZ}

	\author{S.~Dolan}
	\email[Contact e-mail: ]{Stephen.Joseph.Dolan@cern.ch}
	\affiliation{\affiliationCERN}

	\author{D.~Sgalaberna}
    \email[Contact e-mail: ]{davide.sgalaberna@cern.ch}
    \affiliation{\affiliationETHZ}

	\author{A.~Nikolakopoulos}
	\affiliation{\affiliationGhent}

	\author{T.~Dealtry}
	\affiliation{\affiliationLancaster}

	\author{S.~Bolognesi}
	\affiliation{\affiliationCEA}

	\author{L.~Pickering}
	\affiliation{\affiliationRHUL}

	\author{A.~Rubbia}
    \affiliation{\affiliationETHZ}

\begin{abstract}
\noindent
The potential for mis-modeling of \numu/\nue, \numubar/\nuebar and \nue/\nuebar cross section ratios due to nuclear effects is quantified by considering model spread within the full kinematic phase space for CCQE interactions. Its impact is then propagated to simulated experimental configurations based on the Hyper-K and ESS$\nu$SB experiments. Although significant discrepancies between theoretical models is confirmed, it is found that these largely lie in regions of phase space that contribute only a very small portion of the flux integrated cross sections. Overall, a systematic uncertainty on the oscillated flux-averaged \nue/\nuebar cross-section ratio is found to be $\sim$2\% and $\sim$4\% for Hyper-K and ESS$\nu$SB respectively.

\end{abstract}

\maketitle


Currently-running accelerator-based long-baseline (LBL) neutrino experiments,
T2K~\cite{T2K:2019bcf,T2K:2021xwb} and NOvA~\cite{NOvA:2019cyt,NOvA:2021nfi},
are placing increasingly tight constraints on neutrino oscillation parameters.
LBL experiments infer both
(anti)electron neutrino appearance
and (anti)muon neutrino disappearance 
in an (anti)muon neutrino beam
using a ``far'' detector (FD), placed a few hundred kilometres away from the neutrino production point.
LBL measurements are sensitive to the neutrino oscillation parameters:
$\theta_{23}$ (including the octant),
the complex phase $\delta_{CP}$, responsible for the violation of the leptonic Charge-Parity (CP) symmetry, 
and the neutrino mass-squared splittings,
$\Delta m^2_{32}$,
including the neutrino mass ordering (MO), i.e. whether $\Delta m^2_{32} > 0$ (normal) or $\Delta m^2_{32} < 0$ (inverted).
Although the latest LBL measurements 
remain statistically limited,
their sensitivity is continuing to improve as larger samples of data are collected in higher intensity beams~\cite{Abe:2019fux}.
The upcoming Hyper-K~\cite{Hyper-Kamiokande:2018ofw} and DUNE~\cite{Acciarri:2015uup} experiments will 
identify the correct neutrino
MO
and measure $\delta_{CP}$ with a resolution better than $20^{\circ}$.
Another experiment, ESS$\nu$SB, has proposed to further improve the resolution below $8^{\circ}$ ~\cite{Alekou:2022emd}.
With an order of magnitude or more data, 
future experiments are likely to be dominated by 
systematic uncertainties due to the possible mis-modelling of the neutrino-nucleus interaction cross sections
\cite{Alvarez-Ruso:2017oui}. 
Since the predominant sensitivity to $\delta_{CP}$, the MO and the octant stems from an analysis of (anti)electron neutrino appearance event rates at the FD, the uncertainty on the differences between the (anti)muon neutrino cross sections, which can be constrained at a near detector, and the FD-relevant (anti)electron neutrino cross sections, is especially important~\cite{Scott:2020gng,DUNE:2021tad}.
%
%
%
%

%
For interactions where the range of kinematically allowed energy- and momentum-transfer is comparable to lepton mass differences, nuclear processes which determine the cross section may do so differently for different flavours of neutrinos. 
In particular, previous works have investigated differences in the $\parbar{\nu}_{\mu}$ and $\parbar{\nu}_{e}$ cross sections due to the way nuclear effects change the impact of the restriction the lepton mass places on the allowed kinematic phase space~\cite{Martini:2016eec,Ankowski:2017yvm,Nikolakopoulos:2020alk,Nikolakopoulos:2019qcr,Ankowski:2017yvm}. Other works have shown differences due to radiative corrections~\cite{Day:2012gb}, which have currently been assigned a $\sim$2\% systematic uncertainty on the \nue/\nuebar cross-section ratio at energies around one GeV~\cite{T2K:2019ird}, but recent calculations offer prospects for significant reduction~\cite{Tomalak:2021hec,Tomalak:2022xup}.

In this article the impact of nuclear effects on the cross-section ratios
(\numu /\nue,
\numubar /\nuebar,
and
\nue /\nuebar) of charged-current quasi-elastic (CCQE) interactions
are studied.
CCQE interactions on oxygen nuclei (the dominant interaction and target for the T2K, Hyper-K and ESS$\nu$SB experiments)
are investigated across a variety of state-of-the-art and widely used models. Differences in the ratios between oxygen and carbon nuclei are also considered.
A systematic uncertainty
is derived to cover the observed model spread for the Hyper-K (which is also applicable to T2K) and ESS$\nu$SB experiments in the form of two correlated uncertainties on the \numu /\nue and \numubar /\nuebar cross-section ratios, which together imply an uncertainty on
the \nue /\nuebar ratio.



CCQE neutrino interactions are generated with a flat neutrino flux between zero and two GeV on an oxygen target using the NEUT interaction event generator~\cite{Hayato:2021heg}, using either a Local Fermi Gas (LFG) model (with random phase approximation corrections) based on~\cite{Nieves:2011pp, Bourguille:2020bvw}, or a model that uses the plane-wave impulse approximation using the Benhar spectral function (SF), based on~\cite{Benhar:1994hw}. Note that the axial mass parameter $M_A^{QE}$ is set at NEUT's default values of 1.21~GeV for SF and 1.05~GeV for LFG, although an alternative version of SF using 1.03~GeV is also considered. Another alternative version of SF is considered in which Pauli blocking is disabled. 
NUISANCE~\cite{Stowell:2016jfr} is used to process the simulations and to calculate cross sections. The impact of statistical uncertainties was verified to be small~\cite{supp1}.

The NEUT cross-section predictions are compared among each other and to inclusive cross-section calculations using SuSAv2~\cite{Gonzalez-Jimenez:2014eqa} or a Hartree–Fock (HF) model with and without continuum random phase approximation (CRPA) corrections~\cite{Jachowicz:2002rr,Pandey:2014tza}, produced using the hadron tensor tables prepared for their implementations within the GENIE event generator~\cite{Dolan:2021rdd,Dolan:2019bxf,Andreopoulos:2009rq}. In the HF-CRPA case, the distortion of the outgoing nucleon wavefunction (i.e. FSI) can be disabled so the outgoing nucleon is considered a plane wave (PW). In contrast to commonly-used intranuclear cascade FSI, this treatment changes the predicted inclusive cross sections~\cite{Nikolakopoulos:2022qkq}. 
Each calculation is made for an oxygen target, whilst the HF-CRPA model is also considered for carbon. Together, the considered models, summarised in \cref{tab:model-category}, cover a wide range of approaches to 
account
for nuclear effects and represent those most commonly used for neutrino oscillation analyses. They further include model variations with key processes disabled which, whilst not realistic, provide a means to study their role.

\begin{table}[htb]
    \footnotesize
    \centering
    \begin{tabular}{l|c}
    \hline \hline
    \textbf{Model}  & \textbf{Description}  \\
    \hline
    SuSAv2             &  Model from~\cite{Gonzalez-Jimenez:2014eqa} \\
    \hline
    HF              &  Model from~\cite{Jachowicz:2002rr} w/o CRPA corrections\\
    HF-CRPA              &  w/ CRPA corrections\\
    HF-CRPA PW\textsuperscript{\textdagger}            &  w/ CRPA corrections, plane wave nucleon\\
    HF-CRPA C             &  w/ CRPA corrections, carbon target \\
    \hline
    SF                    &  Model from NEUT based on~\cite{Benhar:1994hw} \\
    SF w/o PB\textsuperscript{\textdagger}              &   w/o Pauli blocking \\
    SF $M_A^{QE} 1.03$              &   w/ modified nucleon axial mass \\
    \hline
    LFG               &  Model from NEUT based on ~\cite{Nieves:2011pp} \\
    \hline
   \end{tabular}
    \caption{
    \label{tab:model-category}
    The list of CCQE cross-section models used in this work. All are calculated for an oxygen target, other than HF-CRPA C. Models marked by \textsuperscript{\textdagger} are not realistic but provide a study of disabling certain effects.
    }
\end{table}

\begin{figure}[tb]
\centering
\includegraphics[width=0.236\textwidth]{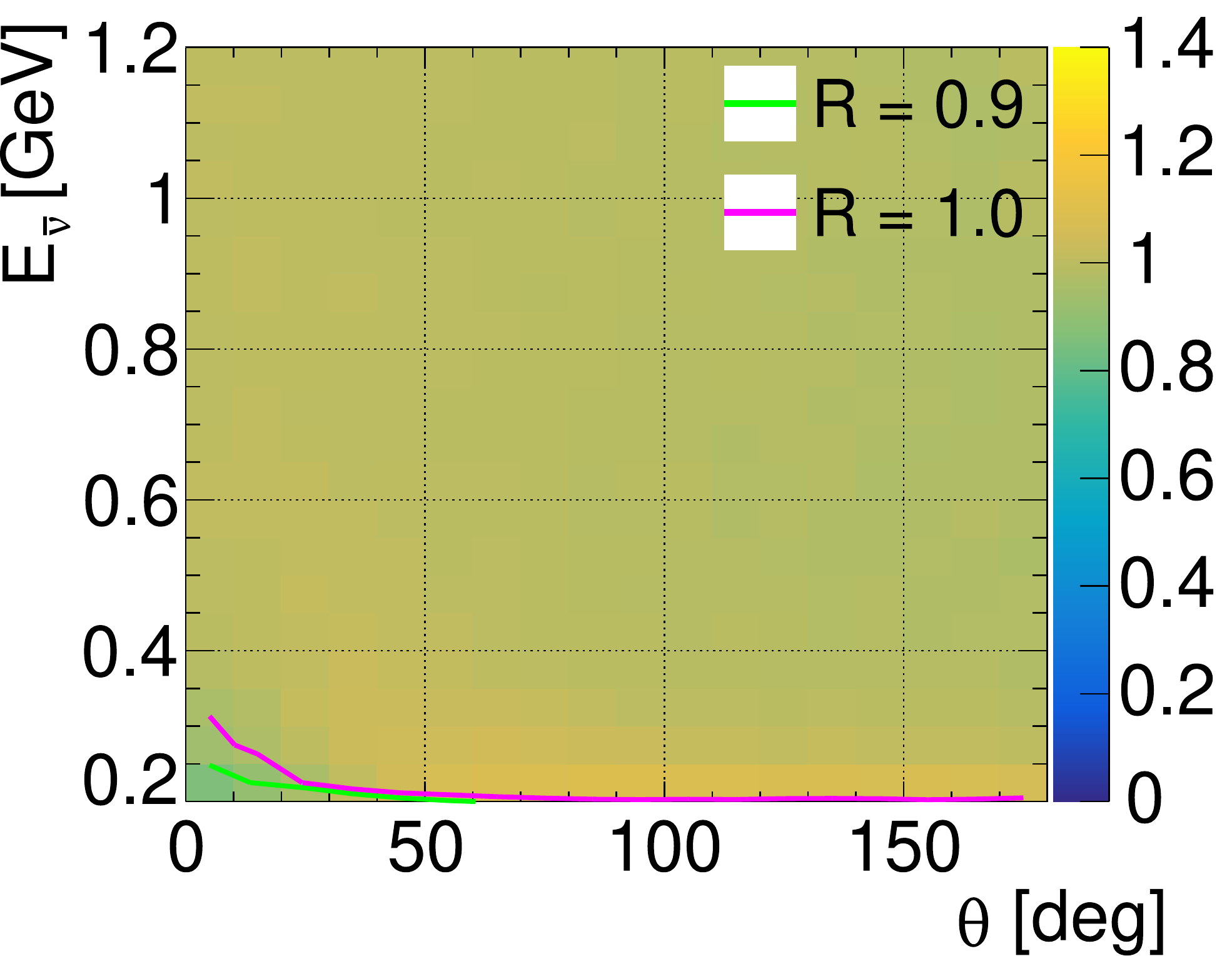}
\includegraphics[width=0.236\textwidth]{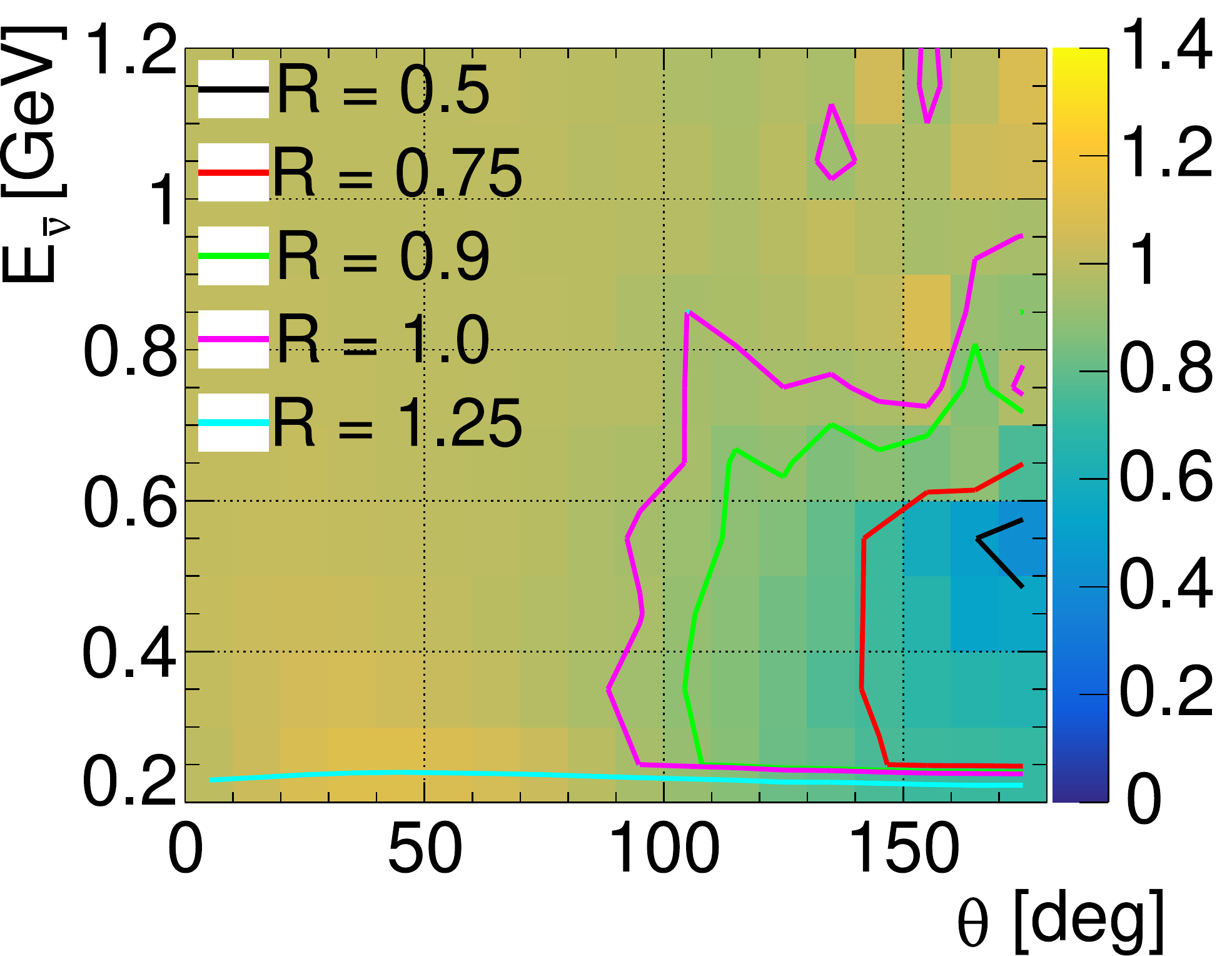}
\includegraphics[width=0.236\textwidth]{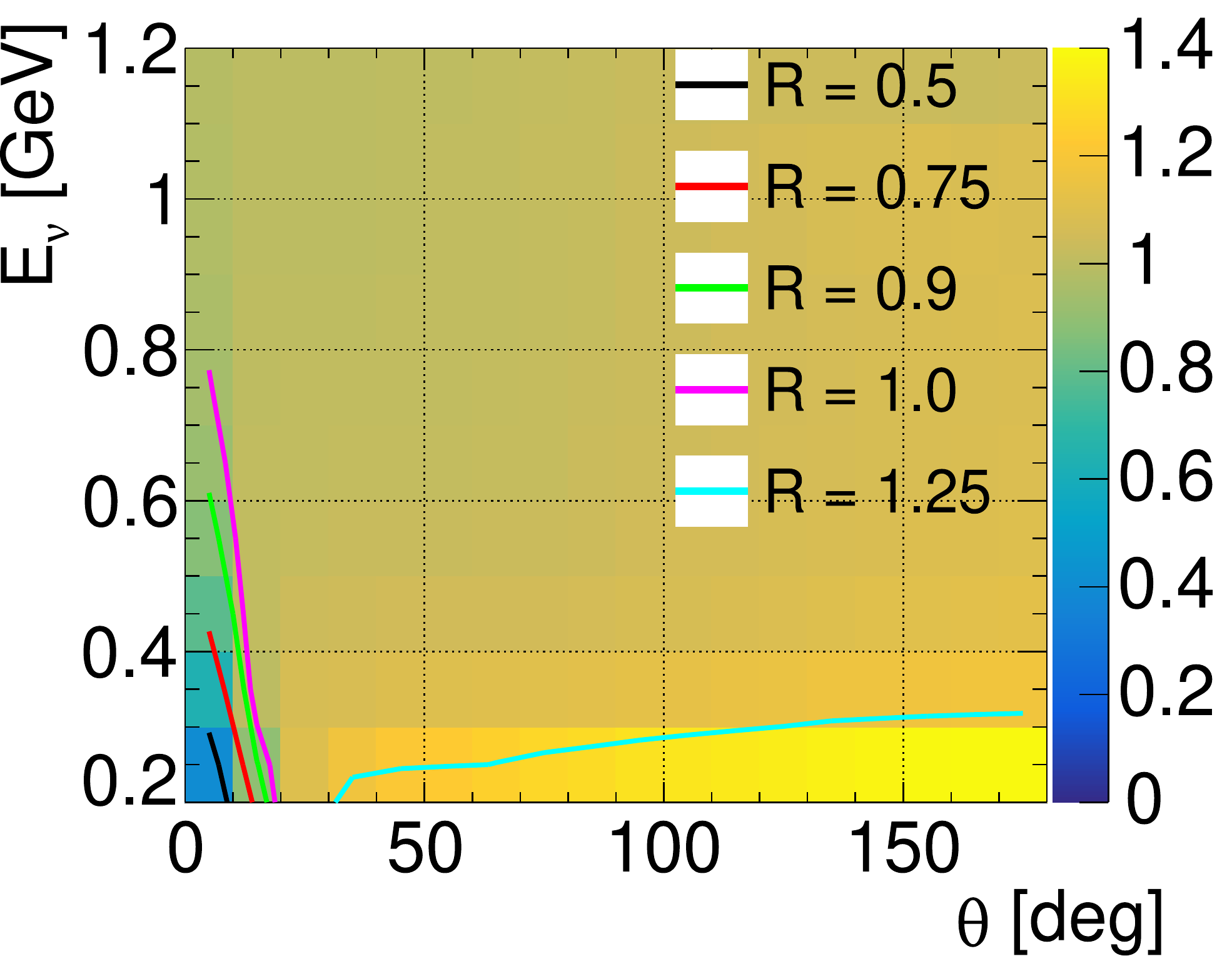}
\includegraphics[width=0.236\textwidth]{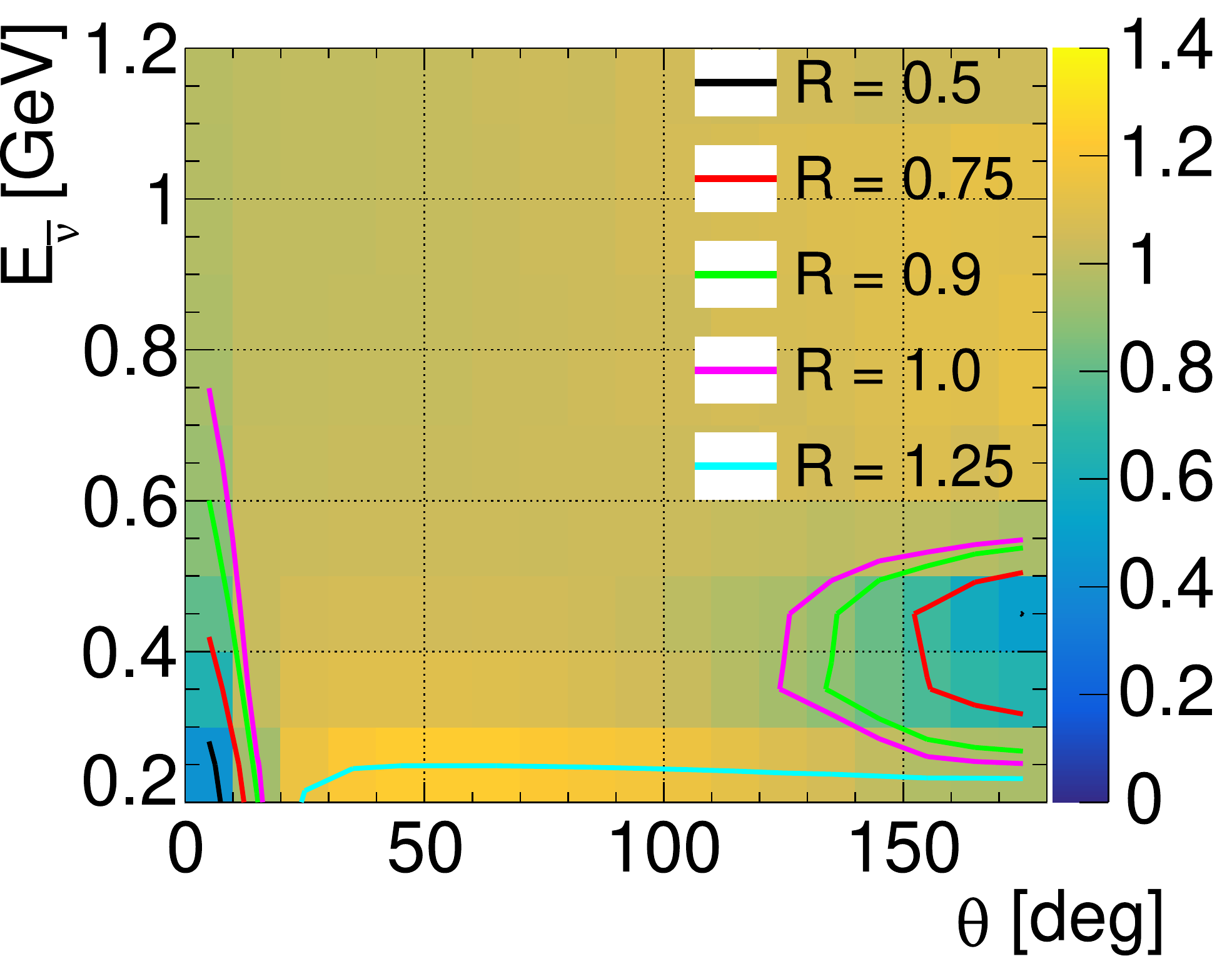}
\caption{
$R_{\nue /\nu_{\mu}}^{\text{SF}}$ (top-left),
$R_{\nue /\nu_{\mu}}^{\text{HF-CRPA}}$ (bottom left),
$R_{\bar{\nu}_e /\bar{\nu}_{\mu}}^{\text{SF}}$ (top-right) 
and
$R_{\bar{\nu}_e /\bar{\nu}_{\mu}}^{\text{HF-CRPA}}$ (bottom-right)
are shown as a function of outgoing lepton angle and the neutrino energy.
The contour lines highlight the regions where the ratio significantly deviates from unity.
}
\label{fig:cross-section-ratio-SF}
\vspace{-5mm}
\end{figure}

A ratio between
$\parbar{\nu}_e$ and $\parbar{\nu}_\mu$ differential cross sections
across a range of
incoming neutrino energy ($E_{\nu}$) and
outgoing lepton angles with respect to the incoming neutrino ($\theta$) is defined as:
\begin{equation*}
R_{\nu_\alpha /\nu_{\beta}}^{\text{Model}} (E_\nu, \theta) =
\left[
\nicefrac{
\frac{d\sigma_{\nu_\alpha} }{d\cos\theta}
}{
\frac{d\sigma_{\nu_{\beta}}}{d\cos\theta}
}
\right]^{\text{Model}} (E_\nu, \theta),
\label{eq:singlerat}
\end{equation*}
where $\alpha$ and $\beta$ give the flavours under consideration.  
$R_{\nue /\nu_{\mu}}^{\text{SF}}$, $R_{\nue /\nu_{\mu}}^{\text{HF-CRPA}}$, $R_{\bar{\nu}_e /\bar{\nu}_{\mu}}^{\text{SF}}$ and $R_{\bar{\nu}_e /\bar{\nu}_{\mu}}^{\text{HF-CRPA}}$ are shown in \cref{fig:cross-section-ratio-SF}\cite{supp4}. 
Note that the contour lines shown are built using a bi-linear interpolation based on the four nearest bin centres~\cite{ROOT} and that this uses unseen bins for $E_\nu<$~2~GeV.
Large differences between the HF-CRPA and SF models are seen in the forward scattered region, as previously studied in \cite{Nikolakopoulos:2019qcr}.
Although this behaviour is also observed in SF, it is much weaker. 

To better quantify these deviations, the double ratio of the differential cross section predicted by two different models is computed as:

\begin{equation*}
    RR_{\nu_\alpha /\nu_{\beta}}^{\text{Model 1/\text{Model 2}}} (E_{\nu},\theta)
    =
    \frac{
    R_{\nu_\alpha /\nu_{\beta}}^{\text{Model 1}}
    (E_{\nu},\theta)
    }{R_{\nu_\alpha /\nu_{\beta}
    }^{\text{Model 2}}
    (E_{\nu},\theta)
    }.
    \label{eq:doublerat}
\end{equation*}

\begin{figure*}[htb]
    \vspace{-5mm}
    \centering
    \includegraphics[width=0.32\textwidth]{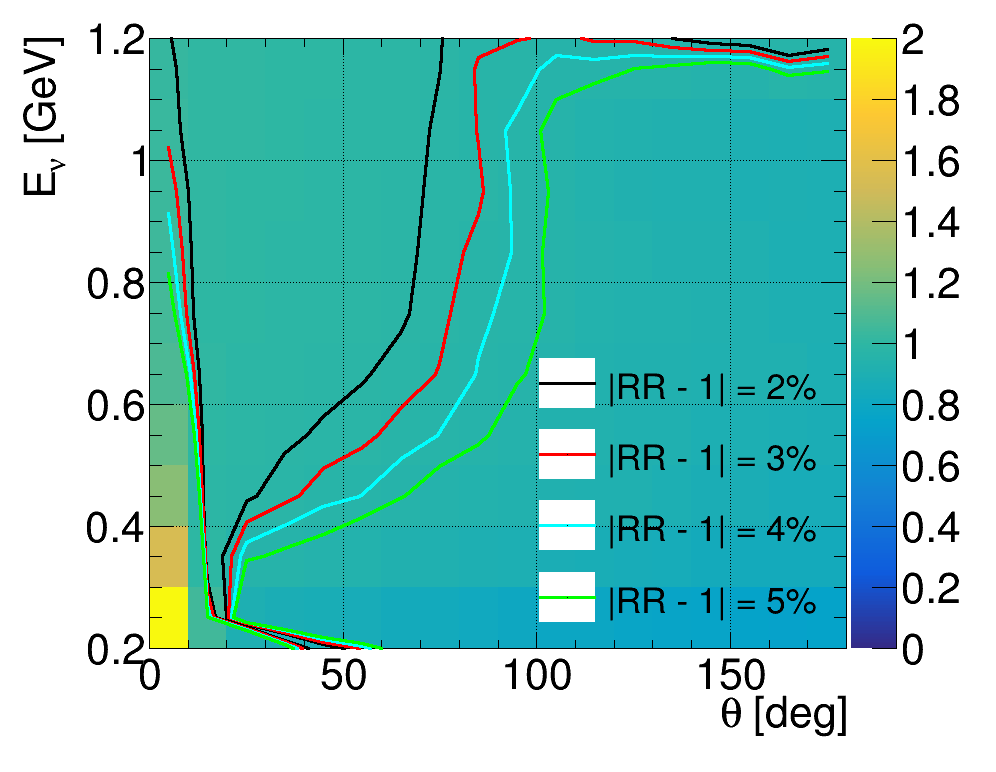}
    \includegraphics[width=0.32\textwidth]{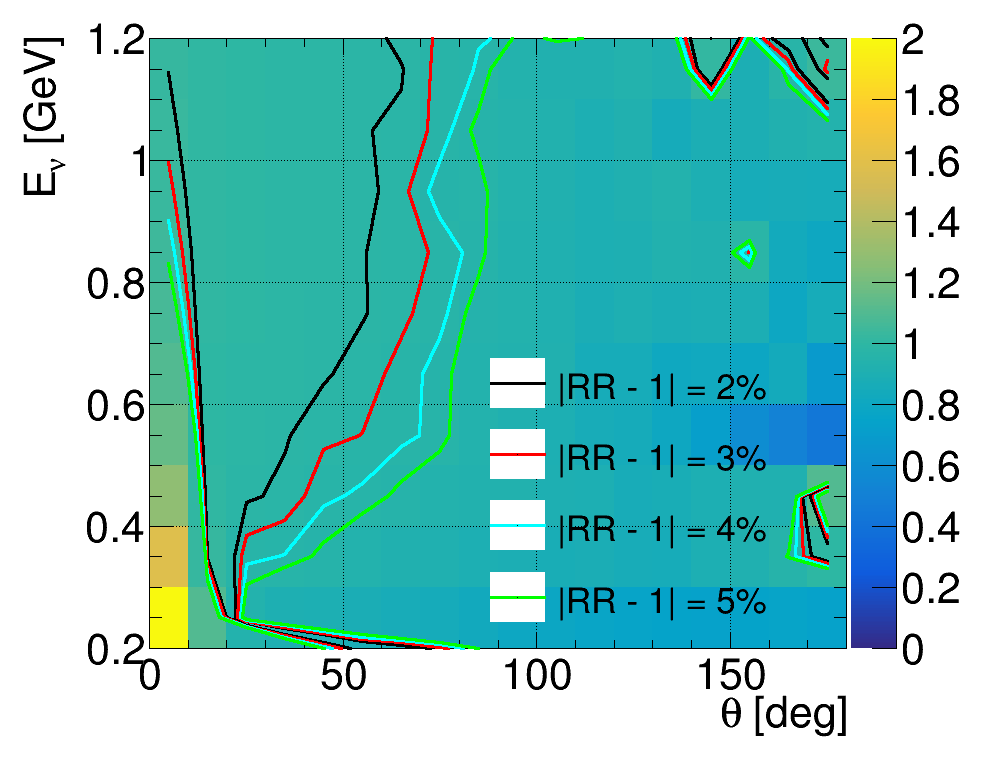}
    \includegraphics[width=0.32\textwidth]{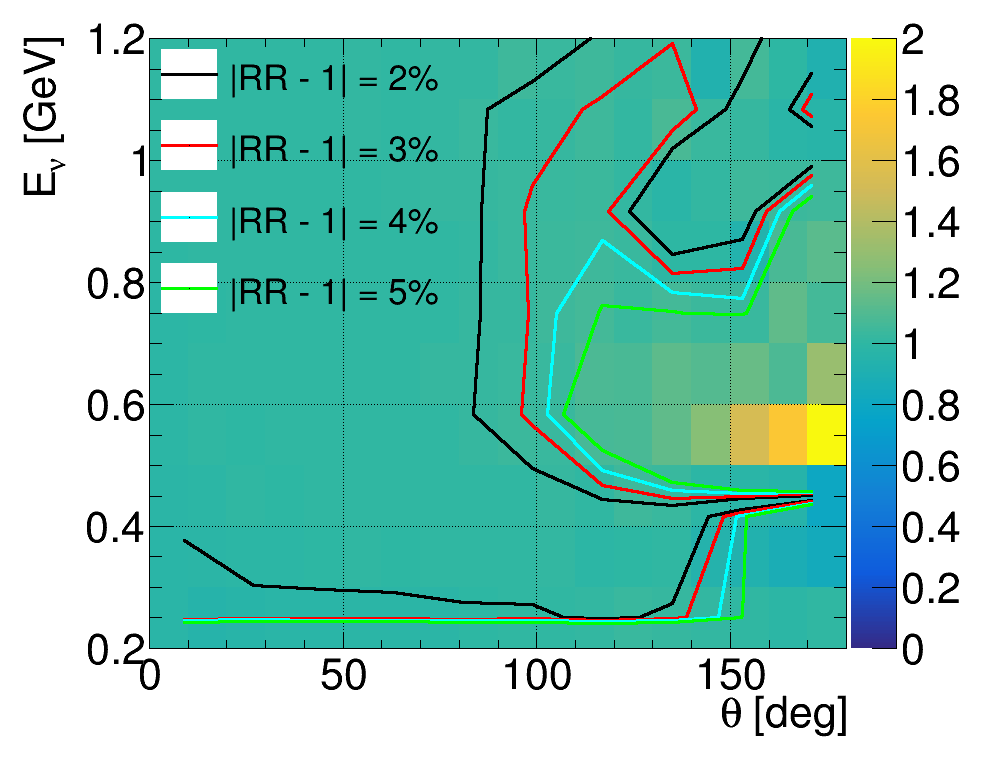}
    \includegraphics[width=0.32\textwidth]{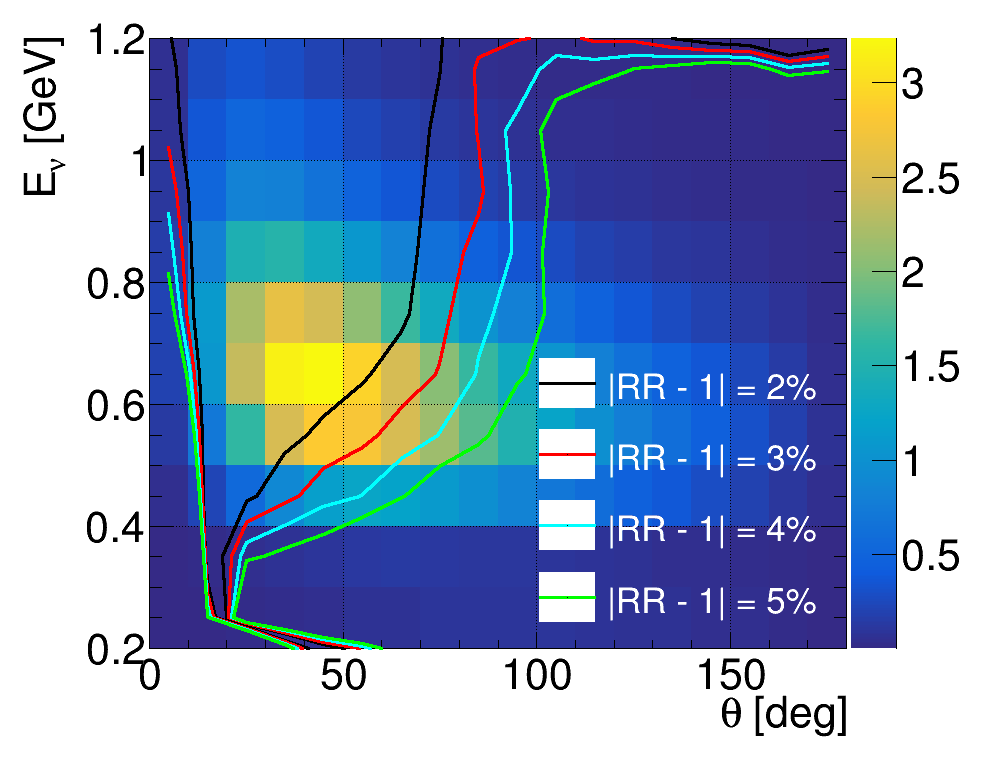}
    \includegraphics[width=0.32\textwidth]{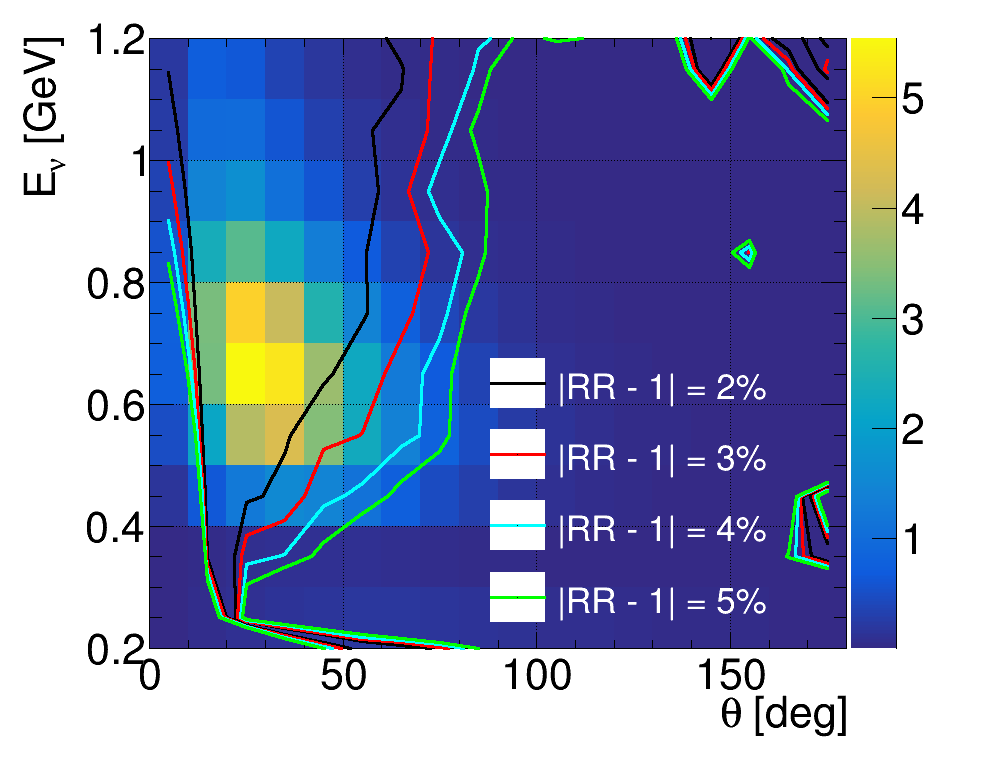}
    \includegraphics[width=0.32\textwidth]{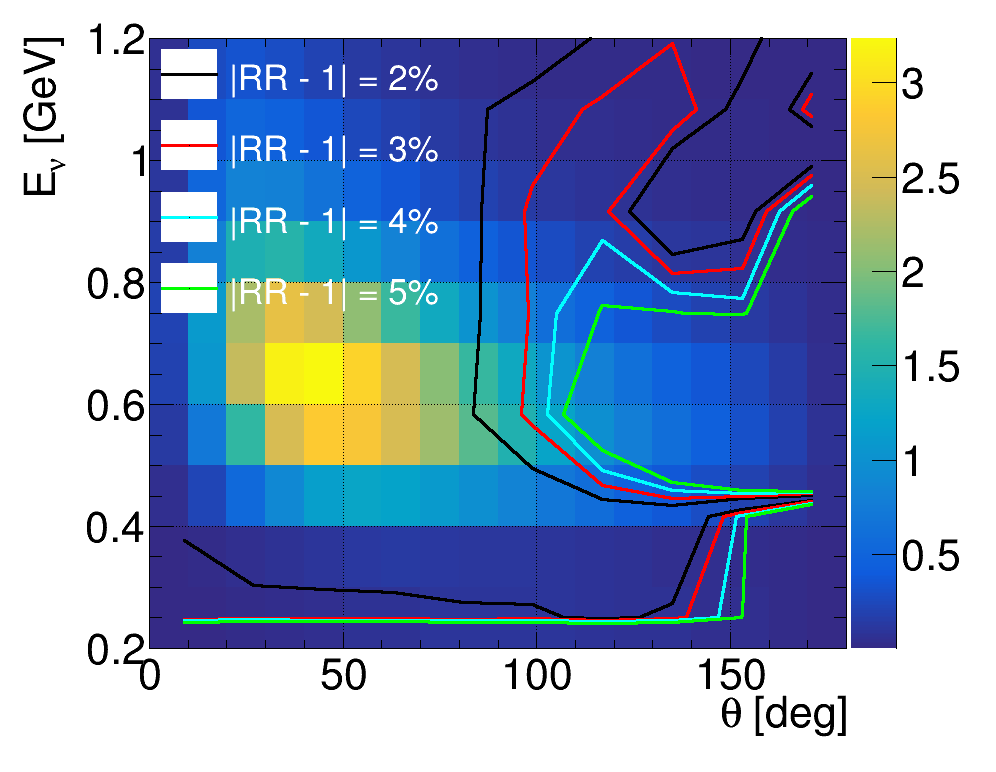}
    \caption{
    The upper plots show $RR_{\nu_e/\nu_{\mu}}^{\text{HF-CRPA/\text{SF}}}$ (left),    $RR_{\bar{\nu}_e/\bar{\nu}_{\mu}}^{\text{HF-CRPA/\text{SF}}}$ (centre) and $RR_{\nu_e / \bar{\nu}_e}^{\text{HF-CRPA/SF}}$ (right) 
    as a function of the outgoing lepton angle and neutrino energy.
    Contour lines highlight regions where $| RR_{\nu_e/\nu_{\mu}} - 1|$ differs from zero. The lower plots show the same contour lines overlaid on the oscillated event rates expected at the T2K/Hyper-K FD built using the SF model (the \nue event rate is shown for the $RR_{\nu_e/\nu_{\mu}}$ and $RR_{\nu_e / \bar{\nu}_e}$ contours and the \nuebar rate is shown for the $RR_{\bar{\nu}_e/\bar{\nu}_{\mu}}$ contours)~\cite{supp3}. The z-axes of the lower plots show the relative proportion of the event rate in each bin as a percentage.
    }
    \vspace{-3mm}
    \label{fig:e_mu}
\end{figure*}

$RR_{\nu_e/\nu_{\mu}}^{\text{HF-CRPA/\text{SF}}}$,    $RR_{\bar{\nu}_e/\bar{\nu}_{\mu}}^{\text{HF-CRPA/\text{SF}}}$ and $RR_{\nu_e / \bar{\nu}_e}^{\text{HF-CRPA/SF}}$,
are shown in \cref{fig:e_mu}. The forward scattered region at angles below 20 degrees
show
a large 
discrepancy
between the models. However, it is interesting to see that the differences remain non-negligible
when considering angles larger than about $50^{\circ}$ 
for energies close to Hyper-K's oscillation maximum
($\sim$ 0.6 GeV). 


In order to investigate the impact of potential cross-section mismodelling,
the contours highlighting the regions with large $RR_{\parbar{\nu}_e/\parbar{\nu}_{\mu}}^{\text{HF-CRPA/SF}}$ are also shown overlaid on expected oscillated $\nue$ and $\nueb$ appearance event distributions at T2K/Hyper-K \footnote{The event rates are calculated using only CCQE interactions (using the SF model), without applying efficiency corrections or detector smearing. The oscillation parameters used are:
 $\sin^2 \theta_{12} = 0.297$,
  $\sin^2 \theta_{13} = 0.0214$,
  $\sin^2 \theta_{23} = 0.526$,
  $\Delta m^2_{21} = 7.37\times 10^{-5}$,
  $|\Delta m^2_{32}|= 2.463\times10^{-3}$,
  $\Delta m_{32}^2>0$ (normal ordering),
  $ \delta_{CP} = 0$.
} in \cref{fig:e_mu}~\cite{supp3}.

From \cref{fig:e_mu} it is clear that neither the large differences in the very forward region, 
nor the differences at low neutrino energies
will have any significant impact on
T2K or Hyper-K
oscillation analyses, as only a very small portion of CCQE interactions will fall within this region. 
However, it can also be seen that a sizeable
fraction of the interactions fall in the higher angle region of the phase space where $RR_{\parbar{\nu}_e/\parbar{\nu}_{\mu}}^{\text{HF-CRPA/SF}}$ differs from unity by more than 2\%. In the case of antineutrino interactions, which have a larger portion of their cross section at more forward outgoing lepton angles, the overlap with regions of large deviations from unity is smaller. The computed $RR_{\nu_e / \bar{\nu}_e}^{\text{HF-CRPA/SF}}$ is also shown, from which it can be seen that the regions with the largest deviations from unity overlap only with the extreme tails of the expected event distribution (i.e. at very low cross section).

An estimate of the integrated uncertainty on the expected $\parbar{\nu}_e$ appearance event rates associated with differences between $\parbar{\nu}_e$ and $\parbar{\nu}_\mu$ cross sections due the modelling of nuclear effects is
computed by
averaging the model differences over the
distribution of events
predicted with the SF model, as illustrated in the lower plots of in \cref{fig:e_mu}.
The resultant uncertainties on 
the \numu /\nue, \numubar /\nuebar, and \nue /\nuebar cross-section ratios are defined respectively as: $\Delta^{\text{Model 1/Model 2}}_{\nue/\num}$, $\Delta^{\text{Model 1/Model 2}}_{\nueb/\numb}$ and $\Delta^{\text{Model 1/Model 2}}_{\nu_e/\overline{\nu}_e}$.
The former two are either fully correlated or fully anti-correlated, depending on whether the averaged model differences cause the cross-section ratios
to change in the same direction.



\begin{figure*}[htb]
 \vspace{-3mm}
\centering
\includegraphics[width = 0.30\textwidth]{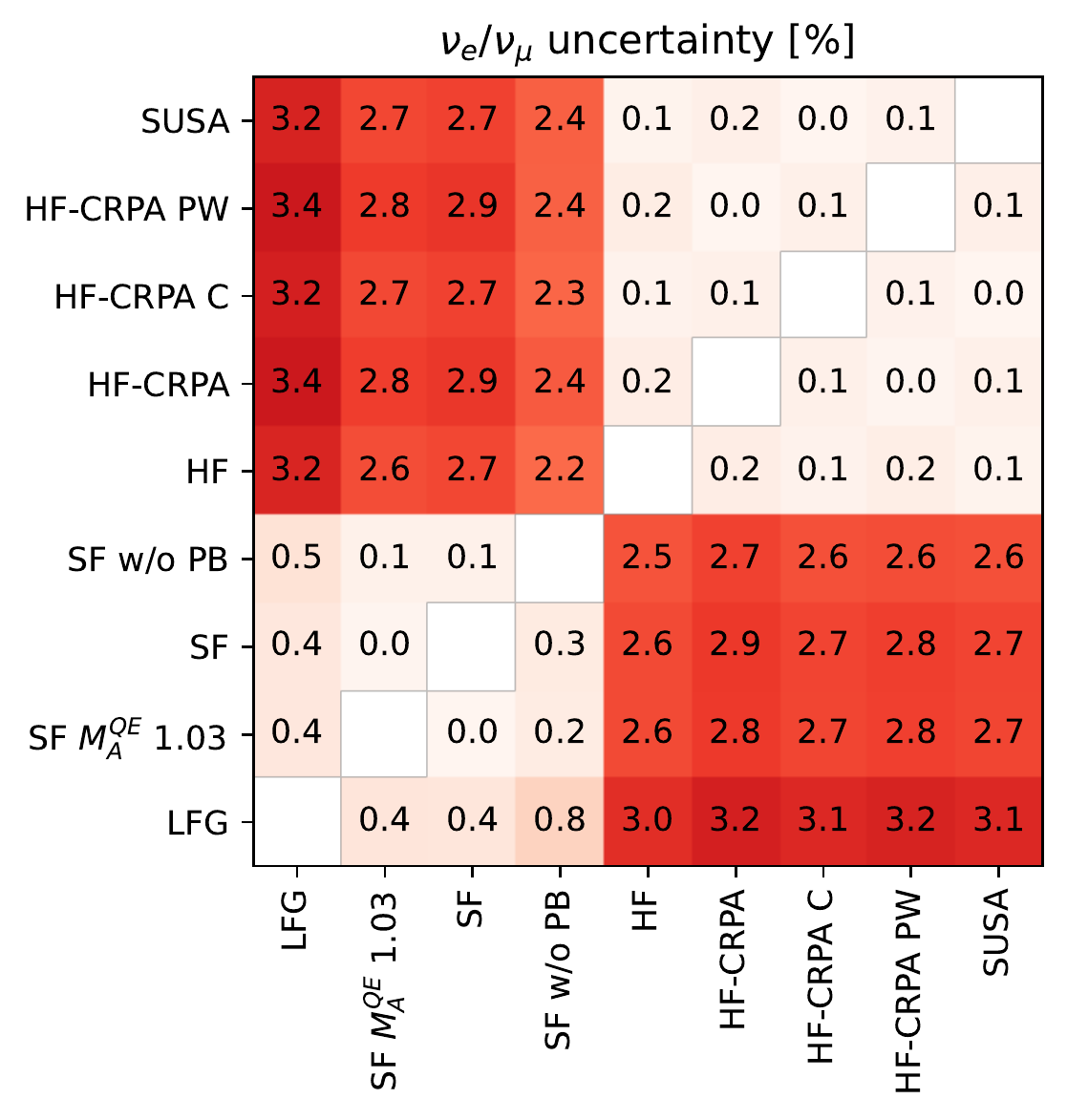}
\includegraphics[width = 0.30\textwidth]{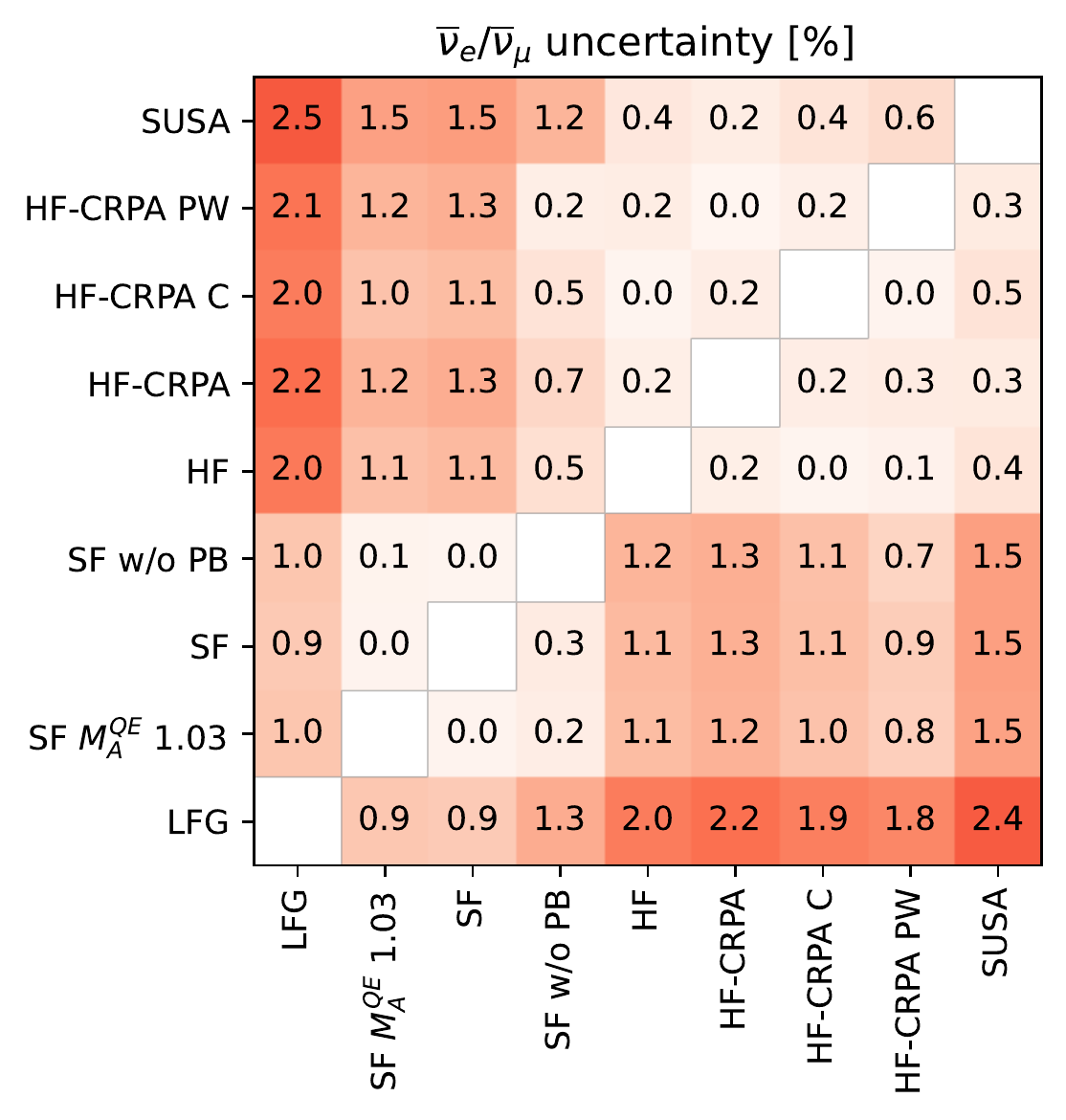}
\includegraphics[width = 0.30\textwidth]{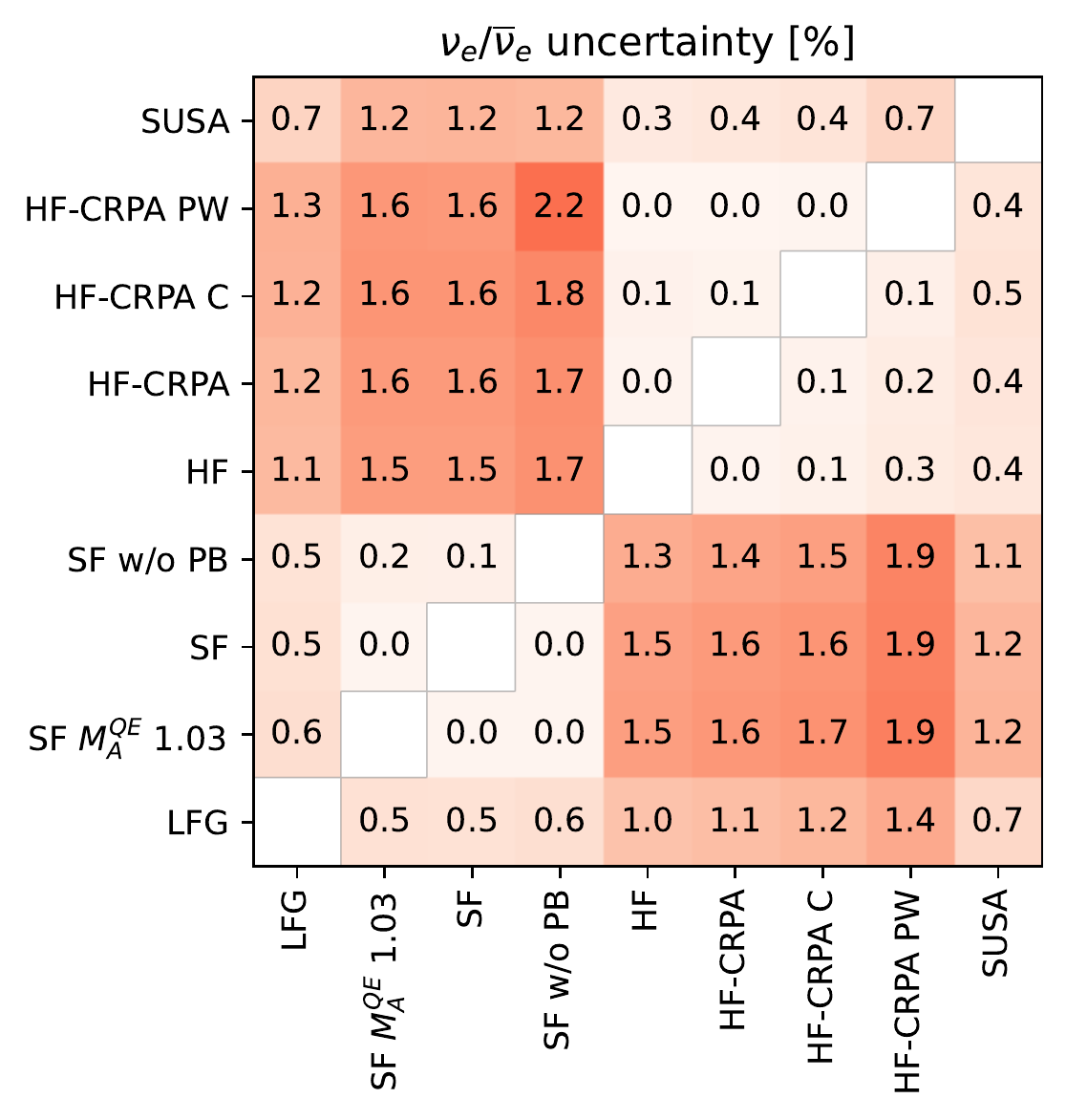}

\caption{The
flux-averaged
uncertainties
in percent obtained by comparing the different cross section models shown in \cref{tab:model-category}: 
$\Delta_{\nu_e / \nu_{\mu}}$ (left),
$\Delta_{\bar{\nu}_e / \bar{\nu}_{\mu}}$ (centre),
$\Delta_{\nu_e / \bar{\nu}_{e}}$ (right).
The lower triangle
is averaged over the event rate distribution predicted by the model given on the horizontal-axis, while the upper triangle contains the resulting values from the averaging over the model on the vertical-axis, resulting in an asymmetric matrix.
}
\vspace{-3mm}
\label{fig:uncert_plot}
\end{figure*}

The flux-averaged uncertainties derived for comparisons of each pair of models introduced in \cref{tab:model-category} are shown as a matrix in \cref{fig:uncert_plot}. This pairwise comparison derived from different model combinations permits an analysis of the possible physical source of differences in predictions of $R_{\nu_e/\nu_\mu}$, $R_{\Bar{\nu_e}/\Bar{\nu_\mu}}$ and  $R_{\bar{\nu}_e/\nu_{e}}$. Overall, every systematic alteration within a model is found to change the ratios of interest by less than 0.5\%, whilst differences between models using different nuclear ground states (LFG, SuSAv2, SF-based, HF-based) are much larger (2-3\%). This may suggest that that the differences are driven by the treatment of the nuclear ground state. Note also that the change in the ratios for HF-CRPA between oxygen and carbon targets is much smaller than the differences between models, implying carbon-to-oxygen differences are likely to be a subdominant effect~\cite{supp2}.  

\begin{figure}[h]
    \vspace{-4mm}
    \centering
    \includegraphics[width = 0.33\textwidth]{t2k.pdf}
    \includegraphics[width = 0.33\textwidth]{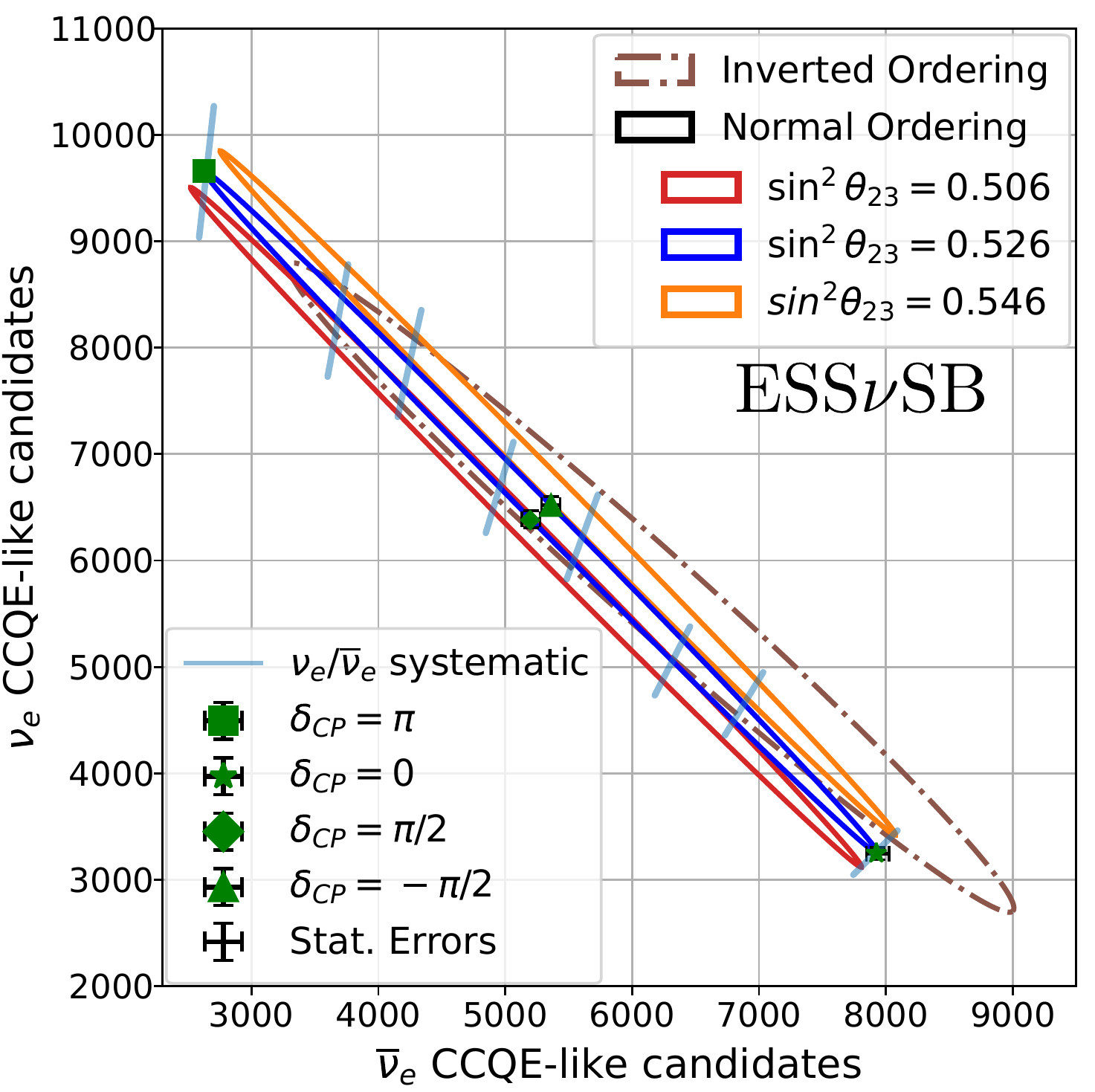}
    \caption{
    Bi-event plots for Hyper-K and ESS$\nu$SB, considering exposures of $2.7 \times 10^{22}$  and $21.6 \times 10^{22}$ protons on target respectively (corresponding to 10 and 1 operational years)~\cite{Hyper-Kamiokande:2018ofw,Alekou:2022emd,supp3}. Each ellipse spans values of $\delta_{CP}$ whilst the different ellipses show variations to the MO and  $\sin^2 \theta_{23}$. For several points around one ellipse, the black bars show the expected experimental statistical uncertainty and the diagonal light blue bar shows the modelling uncertainty from $\Delta_{\nu_e/\bar{\nu}_e}^{HF-CRPA/SF}$.
    }
    \label{fig:Ellipse}
    \vspace{-5mm}
\end{figure}

An indication of the impact of the derived uncertainties on neutrino oscillation analyses can be visualised using ``bi-event'' plots.
These show the expected $\nu_{\mu} \rightarrow \nue$ versus $\bar{\nu}_{\mu} \rightarrow \nueb$ appearance event rate at the FD for different values of the oscillation parameters. Such plots are shown for different values of $\delta_{CP}$, the MO and $\sin^2\theta_{23}$
in~\cref{fig:Ellipse}.
The separation between different oscillation models is compared with the statistical uncertainty and the systematic uncertainty from $\Delta_{\nu_e/\overline{\nu}_e}^{\text{HF-CRPA/SF}}$ (whilst $\Delta_{\nu_e/\overline{\nu}_e}^{\text{HF-CRPA PW/SF w/o PB}}$ shows a larger uncertainty, it is a comparison of two unrealistic models).
The uncertainties are shown to be comparable in size, but the latter is fully correlated between $\nu_e$ and $\overline{\nu}_e$. 
The propagation of the uncertainty through an oscillation analyses therefore
mostly affects the sensitivity to the CP-conserving term (proportional to $\cos \delta_{CP}$) of the oscillation probability, rather than the CP violating one (proportional to $\sin \delta_{CP}$).
It does not extend the range of $\delta_{CP}$ values for which there is degeneracy between the different
MO
and $\delta_{CP}$ but it does enhance the existing significant degeneracy in regions where the ellipses for the different MO overlap.
Fig.~\ref{fig:Ellipse} also shows that a stronger degeneracy is introduced in the measurement of $\sin^2 \theta_{23}$, whose effect is correlated between $\nue$ and $\nueb$ events.
The derived systematic uncertainty
can therefore affect the determination of the $\theta_{23}$ octant.

Similar conclusions can be derived for an ESS$\nu$SB experimental configuration~\cite{supp3}.
The largest
deviation of $RR_{\nu_e / \bar{\nu}_e}$ from unity
was found from the comparison of the
SF and the HF-CRPA models,
resulting in
$\Delta_{\nue/\num} = 6.4\%$,
$\Delta_{\nueb/\numb} = 2.2\%$ and $\Delta_{\nue/\nueb} = 4.2\%$,
considerably larger than the uncertainties derived from the Hyper-K simulation.
Bi-event plots for ESS$\nu$SB are also shown in \cref{fig:Ellipse}.
The impact of the estimated systematic uncertainty
is shown to be
much larger than the projected statistical uncertainties and
significantly impacts
the sensitivity to determining the
$\sin^2 \theta_{23}$ octant.
However, 
note that ESS$\nu$SB gains more from measurements of the shape of the oscillated spectrum which is not reflected in the bi-event plots. 



In conclusion, an evaluation of uncertainties on the $\parbar{\nu}_e$/$\parbar{\nu}_{\mu}$ and $\nu_e$/$\bar{\nu}_e$ cross-section ratios from the modelling of nuclear effects has been studied using the spread of predictions from a wide variety of models. Overall, it has been found that such uncertainties are unlikely to be dominant in measurements of $\sin \delta_{CP}$ term and the MO, although they may become crucial for analyses of $\cos \delta_{CP}$ and the $\sin^2 \theta_{23}$ octant. 
More detailed studies are required in order to evaluate the impact of a systematic uncertainty affecting the modelling of the cross section as a function of FD observables. Whilst this analysis has focused on CCQE interactions, analogous model discrepancies may exist for other processes and nuclei.

The authors would like to thank the T2K and Hyper-K collaborations, in particular for useful discussions in T2K's ``Physics and Performance'' and ``Neutrino Interactions'' working groups. The authors would like to thank Jan Sobczyk, Marco Martini, Claudio Giganti and Anna Ershova for their insightful comments on a draft version of the manuscript. SD would like to especially thank Kevin McFarland, Laura Munteanu and Callum Wilkinson for fruitful discussions. DS and TDi were supported by the Swiss National Science Foundation Eccellenza grant (SNSF PCEFP2\_203261), Switzerland. TDe was supported by the Science and Technology Facilities Council (grant numbers  ST/X002489/1, ST/V006215/1). LP is supported by a Royal Society University Research Fellowship  (URF\textbackslash{}R1\textbackslash{}211661). Fermilab is operated by the Fermi Research Alliance, LLC under contract No.~DE-AC02-07CH11359 with the United States Department of Energy.

\vspace{-2mm}

\bibliography{biblio_asciifree_arxiv}

\end{document}


\title{Supplementary material}
\maketitle

\section{Simulation of experimental configurations}
\label{app:simulation}

The method adopted to simulate and compute the neutrino interaction spectra for is described for both the Hyper-K and ESS$\nu$SB experimental configurations. \\

\paragraph*{\textbf{Hyper-K simulation}:}
An experiment corresponding to ten years of 
data taking with a 187~kt of water target
and an antineutrino beam mode exposure three times larger than the neutrino one, as reported in~\cite{Abe:2018uyc}, is simulated 
with the NEUT neutrino event generator
using the
SF CCQE cross-section model 
and the T2K/Hyper-K 
flux~\cite{Abe:2012av,t2kfluxurl}. 
The Prob3++ software was used to calculate the neutrino oscillation probabilities~\cite{prob3ppurl} and,
for a given set of oscillation parameter values, the number of expected events was computed as:
\begin{equation}
   N_{osc} (E_{\nu}) =  \sum_{E_\nu} 
   \Phi(E_\nu) \times
   \sigma (E_\nu) \times  
    N_{tgts} \times N_{POT} \times 
   P_{osc}(E_\nu)
\end{equation}
where $E_\nu$ is the neutrino energy, $\Phi$ is the incoming neutrino flux (before oscillations occur), $\sigma$ is the interaction cross section,
$N_{tgts}$ is the number of target nuclei,
$N_{POT}$ is the number of accelerator protons on target (POT) 
and $P_{osc}$ 
is the oscillation appearance probability, either
$\nu_{\mu} \rightarrow \nu_e$ for electron neutrino events 
or
$\bar{\nu}_{\mu} \rightarrow \bar{\nu}_e$ for electron antineutrino events. 
Since $N_{osc}$ corresponds to the expected number of 
CCQE events 
without considering detector efficiencies, smearing effects or background, the resulting total number of events 
is slightly different from the 
numbers reported 
in~\cite{Abe:2018uyc}. \\

\paragraph*{\textbf{ESS$\nu$SB simulation}:}
A method analogous to the one adopted for the Hyper-K simulation was used but using the neutrino and antineutrino flux distributions of the ESS$\nu$SB concept experiment. The fiducial active mass was taken to be 200 kt. The same run plan as for Hyper-K was assumed (three times more antineutrino beam than neutrino beam).
The flux at a baseline of 100~km was obtained from the ESS$\nu$SB CDR~\cite{Alekou:2022emd} using a plot-digitizer~\cite{Rohatgi2022}. 
It is then additionally scaled to a FD baseline of 360~km, assuming a radial broadening ($\sim 1/r^2$) of the beam.
The neutrino and antineutrino oscillated CCQE event distributions predicted by the NEUT SF model are shown in~\cref{fig:essnusb_evt}, using the same nominal oscillation parameters as used throughout the article and a baseline of 360 km. 

The peak is at a lower energy and spans a wider range of angles with respect to the Hyper-K case, especially for antineutrinos. Given that there are more events in the region of kinematic phase space shown, in fig. 2 of the main article, to be susceptible to variations of $R_{\nue /\nu_{\mu}}$, it is unsurprising that a larger uncertainty is found for ESS$\nu$SB compared to the Hyper-K experiment.



\begin{figure*}[htb]
    \centering
    \includegraphics[width=0.49\textwidth]{Plots/ESSnuSB_Flux/h_nu.png}
       \includegraphics[width=0.49\textwidth]{Plots/ESSnuSB_Flux/h_anu.png}
    \caption{
      The digitised ESS$\nu$SB oscillated NEUT SF-based event rates for $\nu_\mu$ (left) and $\overline{\nu}_\mu$(right) expected at the far detector. The X and Y axes show respectively the outgoing lepton angle and the neutrino energy. The Z-axis shows the relative proportion of the event rate in each bin as a percentage. }
    \label{fig:essnusb_evt}
\end{figure*}

\section{Statistical Uncertainties} 
\label{app:statuncert}
\begin{figure*}[htb]
    \centering
    \vspace{-10mm}
    \includegraphics[width=0.90\textwidth]{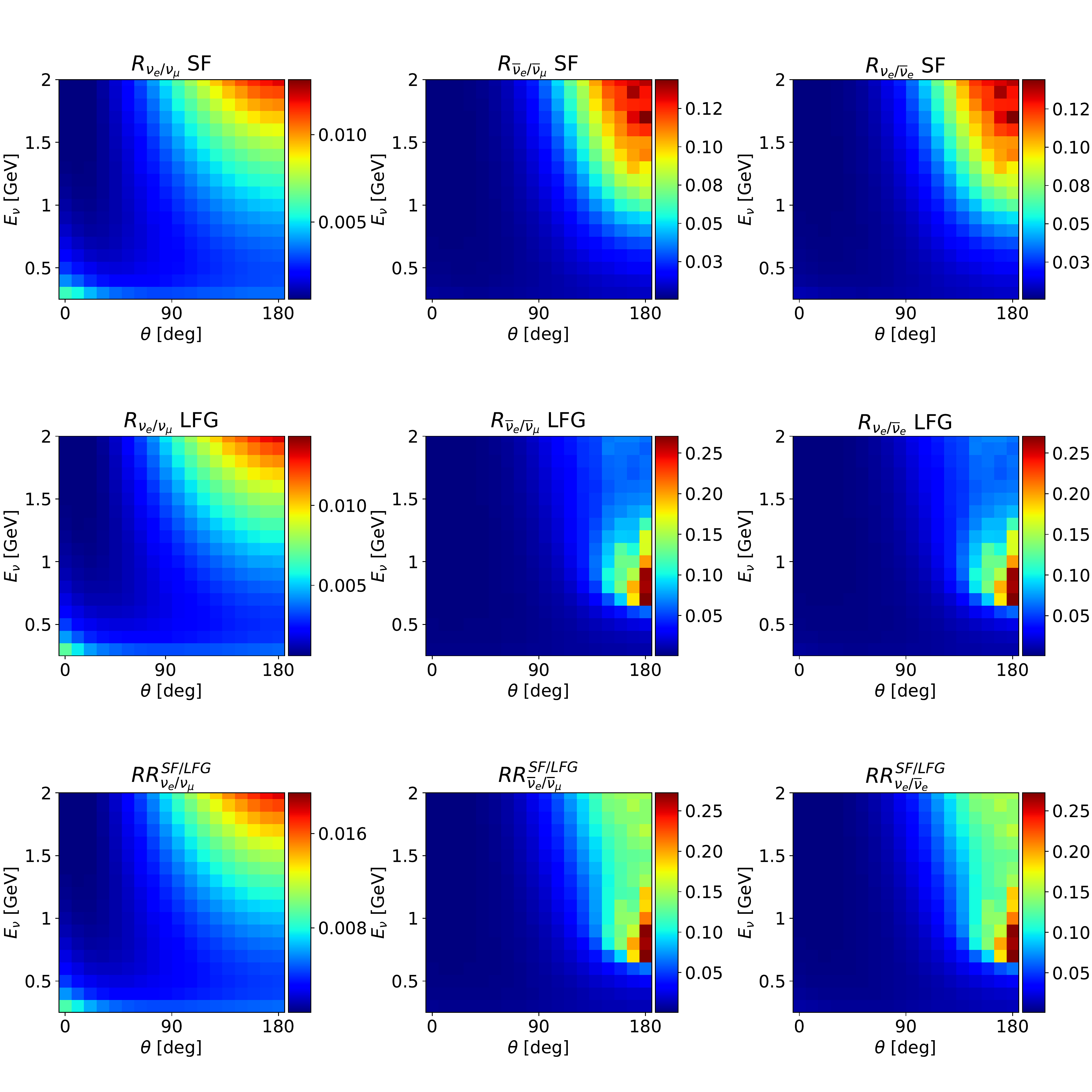}
    \caption{
    The relative fractional statistical uncertainty on 
    single and double cross-section ratios considered 
    for the SF and LFG models.
    }
    \label{fig:stats}
\end{figure*}

The SF and LFG model predictions where extracted using a NEUT Monte-Carlo simulation and are thus subject to a statistical uncertainty related to the number of events generated. For this analysis, 300 million events where produced per neutrino flavour and model. However, even with large numbers of events, the statistical uncertainty on cross-section ratios (or double ratios) is important to consider. The relative statistical uncertainty on a cross-section ratio can be calculated as:

\begin{equation}
\delta(E_\nu,\theta)/\left(N_A/N_B\right) = \sqrt{1/N_A(E_\nu,\theta) + 1/N_B(E_\nu,\theta)}, 
\end{equation}
where $N_A$ and $N_B$ are the number of events within some bin of $E_\nu$, $\theta$ of one of the flavours. Fig.~\ref{fig:stats} shows the relative statistical uncertainty on the cross-section ratios considered within this article, using the two dimensional binning shown in the majority of figures (10 degree bins in $\theta$ and 100 MeV bins in $E_\nu$). Whilst the statistical uncertainty within individual bins shown can be up to $\sim$10\%, it is usually at the $\sim$0-3\% level in the regions of interest for this article.

When double ratios (RR) are considered (i.e. in the ratio of model predictions of two cross section ratios) the statistical uncertainty enters four times when comparing two NEUT-generated models (note that the calculations from the hadron tensor tables are analytical and have no statistical uncertainty). The size of the relative uncertainty on the double ratios of NEUT cross sections are also shown in \cref{fig:stats}, reaching $\sim$20\% for some bins at very high energies and angles but is generally less than $\sim$5\% in regions of interest for this article. 

Overall, it must be noted that the relative statistical uncertainly on the systematic uncertainties calculated within this article, which considers the entire integrated phase space, will be much smaller than the uncertainty on individual bins.

\section{Investigating the source of the nuclear model uncertainties}

The cause of differences in model predictions of $\parbar{\nu}_e$/$\parbar{\nu}_{\mu}$ or $\nu_e$/$\bar{\nu}_e$ cross-section ratios has previously been assigned to a number of physical processes. The calculation of associated model-spread based uncertainty metrics ($\Delta_{\nu_e/\nu_\mu}$, $\Delta_{\overline{\nu}_e/\overline{\nu}_\mu}$, $\Delta_{\nu_e/\overline{\nu}_e}$) for cross-section integrated over the T2K/Hyper-K flux is performed using sets of model that contain systematically different physics in an attempt to establish underlying sources of the uncertainty. The results are shown in fig.~3 of the main article, considering the full list of model shown in tab.~I. This section contains a more detailed breakdown of the impact of each model variation.
\\

\paragraph*{\textbf{Influence of $M_{A}^{QE}$:}}
The nominal values of $M_{A}^{QE}$ in the HF-CRPA and SF models are 1.03 and 1.21 GeV respectively. The impact of changing $M_{A}^{QE}$ to 1.03 GeV in SF, broadly spanning the range of values implied by cross-section measurements~\cite{Wilkinson:2016wmz}, is investigated as a source of uncertainty. The uncertainty metrics used throughout the article are calculated between SF with the two different values of $M_{A}^{QE}$. The resulting uncertainties are negligible (less than 0.1\%) . 
\\ 

\paragraph*{\textbf{Pauli blocking:}}
Pauli blocking introduces a region at small angles and low energies where, counter intuitively, the muon neutrino cross section becomes larger 
than the electron neutrino cross section, as discussed in ~\cite{Nikolakopoulos:2019qcr,Ankowski:2017yvm}. 
In order to investigate the influence of Pauli blocking 
the nominal SF model is compared to a version in which the Pauli blocking was disabled. 

A comparison of the cross section predicted by NEUT's SF model with and without Pauli blocking is shown in \cref{fig:SF_woPB}.
It is clear that the only region of the phase space affected by Pauli blocking is at low neutrino energies and forward lepton angle, where the $\nu_{\mu}$ ($\bar{\nu}_{\mu}$) cross section becomes lower than the $\nu_{e}$ ($\bar{\nu}_{e}$) one.
This region is also affected by a relatively large deviation from unity in $RR^{\text{SF / (SF w/o PB)}}_{\nu_e / \nu_{\mu}}$.
However, this is also a region of very low cross section in all models and so barely overlaps with the distribution of events expected at the Hyper-K or ESS$\nu$SB FDs.

The resulting systematic uncertainties for all metrics considered due to turning Pauli blocking on and off in the SF model is at the level of 0.2\% or less. This small impact is because Pauli blocking affects a very narrow region of kinematic phase space which contains only a very small portion of the total cross section. 
\\

\begin{figure*}[htb]
\centering
\includegraphics[width=.32\linewidth]{Plots/XSec/e_mu_SF_A.png}
    \includegraphics[width=.32\linewidth]{Plots/XSec/e_mu_SF_B.png}
    \includegraphics[width=0.32\linewidth]{Plots/XSec/e_mu_SF_A_B.png}
 \caption{
$R^{\text{SF}}_{\nu_e / \nu_{\mu}}$ (left),
$R^{\text{SF w/o PB}}_{\nu_e / \nu_{\mu}}$ (centre)
and
$RR^{\text{SF / (SF w/o PB)}}_{\nu_e / \nu_{\mu}}$ (right) 
are shown.
The double ratio of those two first plots is shown on the right. The contour lines highlight the regions where the single ratio significantly deviates from unity and are built using a bi-linear interpolation based on the four nearest bin centres~\cite{ROOT}. Note that this uses unseen bins for neutrino energies less than 0.2 GeV.
}
\label{fig:SF_woPB}
\end{figure*}

\paragraph*{\textbf{Inclusion of CRPA nucleon correlations:}}
The HF-CRPA model is based on a mean-field HF approach with CRPA corrections to model long-range nuclear correlations. 
The influence of these corrections on the derived uncertainties is calculated as in the previous paragraphs. 
Overall their inclusion leads to an uncertainty of less than 0.2\% for all metrics.
\\

\paragraph*{\textbf{Final State Interactions:}}
A major difference between the HF-CRPA and SF models is the inclusion of FSI in the calculation of the inclusive cross section via the consideration of the distortion of the outgoing nucleon wavefunction. 
This was identified as part of the cause of the HF-CRPA forward scattering features shown in the main article and discussed in~\cite{Nikolakopoulos:2019qcr}. 
The impact of such FSI was studied within the HF-CRPA model 
by turning its effects on and off.
The derived uncertainty was found to be less than 0.3\%, for all metrics. 
This is due to the small overlap between the heavily impacted forward lepton region 
and the phase space covered by the electron (anti)neutrino appearance events expected at the Hyper-K (or T2K) FD. 
\\

\paragraph*{\textbf{Nuclear model:}}
By far the largest differences in model predictions of $\parbar{\nu}_e$/$\parbar{\nu}_{\mu}$ or $\nu_e$/$\bar{\nu}_e$ cross-section ratios is found when comparing any HF-based or RMF-based model to any SF-based or LFG-based model. Given that the effect of changing any individual component of the models was found to be small,  the difference in the ratios is likely driven by the treatment of the nuclear ground state. Whilst the difference in the cross-section ratios between models is dominated by the very forward and backward regions, the impact on 
Hyper-K 
oscillated $\nu_e$ and $\bar{\nu}_e$
event rates remains driven by the smaller differences at intermediate scattering angles. 
\\

\section{Investigating the 
uncertainty related to differences between carbon and oxygen nuclear models}

It is interesting to consider the sensitivity of the model predictions of $\parbar{\nu}_e$/$\parbar{\nu}_{\mu}$ or $\nu_e$/$\bar{\nu}_e$ cross-section ratios to the target nucleus considered, particularly for the case of differences between oxygen and carbon target nuclei (as these are the predominant nuclear target in T2K/Hyper-K's near detector and far detector respectively).
In order to evaluate this, 
the cross-section calculation 
from HF-CRPA on carbon nucleus was compared with cross-section calculations using an oxygen nucleus. The differences in the cross-section ratios on a carbon and oxygen target are found to be small compared to differences between models on a single target. This suggests that, at least for the HF-CRPA model considered, that carbon-to-oxygen differences appear to have a subdominant effect on the cross-section ratios of interest.

\bibliography{biblio_asciifree}